\documentclass[twocolumn,pra]{revtex4-1}

\usepackage{graphicx}
\usepackage{dcolumn}
\usepackage{bm}
\usepackage{physics}
\usepackage{textcomp}
\usepackage{gensymb}
\usepackage{amsmath}
\usepackage{amssymb}

\newcommand{\kp}{\mbox{$\bm{k}\!\cdot\!\bm{p}$} }

\def \dya #1#2#3#4 {\left|#1\right\rangle_{#2}{\vphantom{\left|#3\right\rangle}}_{2}\hspace{-0.07cm} \left\langle#4\right|}

\begin{document}

\title{Spin-orbit-induced hole spin relaxation in a quantum dot molecule: the effect of
  $s$-$p$ coupling} 

\author{Karol Kawa}
 \email{Karol.Kawa@pwr.edu.pl}
\author{Pawe{\l} Machnikowski}%
\affiliation{Department of Theoretical Physics, Faculty of Fundamental Problems of Technology,
 Wroclaw University of Science and Technology, 50-370 Wroc{\l}aw, Poland}

\begin{abstract}
We study the effect of the coupling between the hole $s$ shell of one quantum dot and the $p$
shell in the other dot forming a quantum dot molecule on the spin relaxation between the
sublevels of the hole $s$ state. Using an effective model that captures
the spin-orbit effects in the $p$ shell irrespective of their origin, we show that the
strong spin mixing in the $p$ shell can be transferred to the $s$ shell of the other dot,
leading to enhanced spin relaxation in a certain energy range around the $s$-$p$ resonance
if the dots are misaligned and the magnetic field is tilted from the sample plane. 
\end{abstract}

\keywords{double quantum dot, spin relaxation, piezoelectric field, phonon bath}
\maketitle

\section{\label{sec:Intro}Introduction}

The ability to confine a single carrier in a semiconductor quantum dot (QD) not only
opened an exciting field of studies but also  paved the way to many possible
applications. The idea of implementing quantum computing schemes on such a paradigmatic
physical qubit \cite{loss98} has driven physical interest for more than 20 years,
leading to the development of sophisticated schemes for spin initialization
\cite{Brash2015,Ardelt2015,Mar2016,Atature2006b,Mar2014}, 
control \cite{press08,Press_NP10,DeGreve2011,Godden2012a},
and readout \cite{Vamivakas2010,Delteil2014}. Coupling and entangling spins to phonons 
\cite{DeGreve2012,Gao2012,Schaibley2013,Vora2015} makes it possible to transfer quantum
states from spins to photons \cite{He2017}, 
entangle distant spins \cite{Delteil2014}, and build quantum repeaters \cite{McMahon2015},
which provides building blocks for quantum networks.

The possible applications of confined spins motivate studies of spin dynamic and, in
particular, spin relaxation in QDs. While phase coherence of the spin may be limited
by charge noise \cite{Godden2012,jovanov11,Prechtel2015}
and fluctuations of the effective
magnetic field due to nuclear spins \cite{merkulov02}, spin relaxation at moderate
magnetic fields is mostly due to
spi-orbit (SO) coupling that enables phonon-mediated spin-flip transitions
\cite{heiss07,bulaev05a}. In contrast to unstrained or uniformly strained systems, where
the SO effects are characterized by usual Rashba \cite{Bychkov1984} or Dresselhaus
\cite{Dresselhaus1955} couplings, in strained self-assembled QDs the situation is more
complex and the SO effect that predominantly affects spin relaxation is induced by
shear strain \cite{Mielnik-Pyszczorski2018a,Gawarecki2018a-pre}.
While the measured hole spin relaxation times vary from experiment to experiment, the observed
upper limit seems to reach hundreds of microseconds
\cite{Dahbashi2012,heiss07,Climente2013,Segarra2015b,Fras2012}. These hole spin life
times are shorter than those for electrons because of the much stronger SO
coupling in the valence band.

It has been shown that systems build of coupled QDs, referred to as quantum dot molecules
(QDMs), offer extended feasibility of spin initialization, manipulation, and readout, due to the
sensitivity of confined states to external electric field and the possibility of exploiting
tunneling processes \cite{Vamivakas2010,muller12b,economou12}. In a QDM, spin-orbit
effects are enhanced by symmetry breaking due to QD off-axial misalignment, which 
typically appears in these structures \cite{Doty2010}. Hole spin flip processes in
QDMs were studied within the \kp method for a simplified QD geometry without strain
\cite{Segarra2015} near the resonance between the $s$ shells in the two QDs. It was shown
that the spin relaxation rate is considerably enhanced by QD misalignment and by
the presence of Dresselhaus SO interaction, with the spin life times around
a $\mu$s at the magnetic field of $2$~T, roughly constant across the
resonance. In a subsequent study \cite{Stavrou2018} the spin-flip rate was also studied for an
electron, using a more realistic model including strain in a QDM built of two nominally
identical QDs.   

The SO coupling that underlies the hole spin relaxation manifests itself in the
band structure of a semiconductor in a rather complicated way that is captured e.g. by various
inter-band terms of the \kp model \cite{winkler03,Planelles2013}. For a carrier confined
in a QD the SO coupling is particularly pronounced in the hole $p$ shell, where it
dominates over the Zeeman terms, leading to reordering of the $p$-shell hole states
\cite{Gawarecki2018}.  It
turns out that the results of 
exact \kp calculations are reproduced by an effective model with a few fitting parameters
determined by comparison to \kp results
\cite{Doty2010,Gawarecki2018}. When supplemented with a simple model of wave functions and
a standard description of 
carrier-phonon couplings, such a model yields a simple, yet reasonably accurate,
description of charge and spin relaxation in a QD, accounting for all the SO effects,
including those induced by strain \cite{Gawarecki2018a-pre}.
Similar effective models were used in the description of
tunneling between QDs in a QDM \cite{stavrou05}.

In this paper we 
study hole-spin relaxation between Zeeman sublevels of the ground ($s$-shell) state in one
of the QDs forming a QDM, resulting from tunnel coupling to the $p$-shell states of the
other QD. We use the effective model for SO couplings in the valence band
extended by tunneling terms along with the standard model of carrier-phonon coupling. We show
that the strong SO coupling in the hole $p$ shell can be transferred via tunnel coupling to
the $s$-shell of the other QD in an axially misaligned QDM leading to considerably
enhanced spin relaxation 
in a tilted magnetic field. The hole-spin flip rate becomes particularly large in the
vicinity of the inter-dot $s$-$p$ resonance but the effect persists also away from the
resonance, where the states are nearly completely localized in one of the QDs and nearly
ideally spin-polarized.

The organization of this paper is as follows.
In Sec.~\ref{sec:Model} we describe the system and introduce the effective
Hamiltonian. In Sec.~\ref{sec:results} we present the results for the  relaxation rates as a
function of electric and magnetic field. Sec.~\ref{sec:concl} concludes the paper. 

\section{\label{sec:Model}Model}
\subsection{Description of the system}
\label{subsec:description_of_the_system}

\begin{figure}[tb]
    \centering
    \vspace{-0.7cm}
    \includegraphics[width=0.7\linewidth]{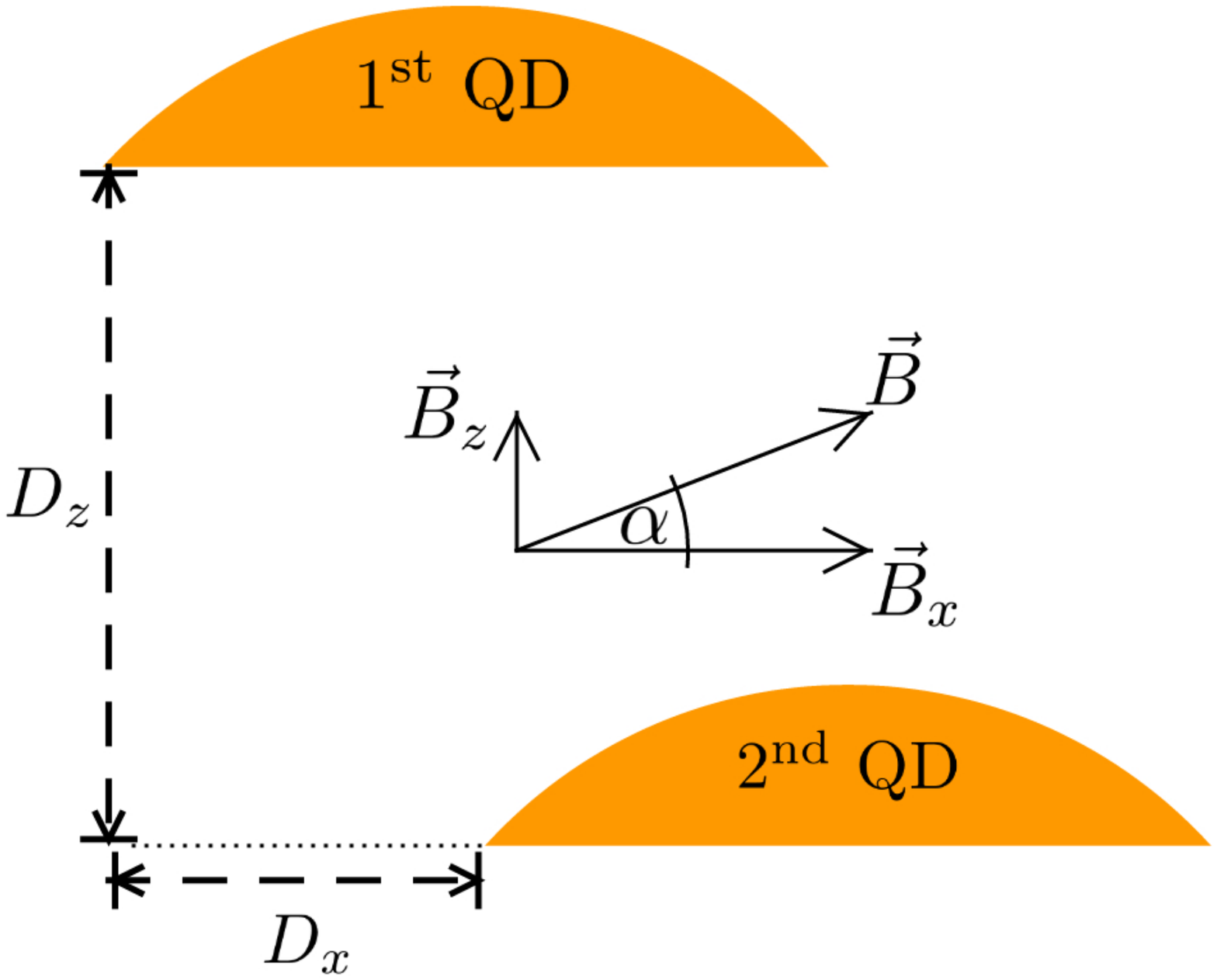}
    \vspace{-1.3cm}\caption{Schematic representation of the system: two QDs stacked along
      the $z$ axis (growth direction) and misaligned with respect to the vertical axis.} 
    \label{fig:QDs}
\end{figure}

The structure consists of two self-assembled lens-shaped QDs grown in the [001]
direction. The dots are stacked vertically with the inter-dot distance $D_z$. The
cylindrical symmetry of the system is broken by a small relative in-plane displacement
$D_x$ (see Fig.~\ref{fig:QDs}). 

The wave function of the hole confined in a QD is the solution of a 2-dimensional quantum
oscillator i.e. Fock-Darwin  state \cite{fock28,darwin30}, with the principal quantum
number $n$ and the angular quantum number $m$ representing the envelope angular
momentum. The wave functions are 
localized in the plane on a distance $l_\bot$ and in the growth direction on a distance
$l_z$.  We take into account the $s$-shell states ($n=0$, $m=0$) in the top QD, whereas in
the bottom QD both the $s$-shell and the $p$-shell states ($n=0$, $m=0$ and $n=1$, $m=\pm
1$) are 
included.  We take into account the two values of the band angular momentum (``spin'') of
a heavy hole, represented by $s=\uparrow,\downarrow$. 
The basis states are then $\{\ket{l,m,s}\} = $ $\{\ket{1, 0 \uparrow}$, $\ket{1, 0
  \downarrow}$, $\ket{2, 0 \uparrow}$, $\ket{2, 0 \downarrow}$, $\ket{2, +1 \uparrow}$,
$\ket{2, +1 \downarrow}$, $\ket{2, -1 \uparrow}$, $\ket{2,-1 \downarrow}\}$, where the
first number $l=1,2$ refers to QD1 and QD2, respectively. 
The off-axis misalignment leads to inter-dot $s$-$p$ coupling because of the
symmetry breaking \cite{Gawarecki2012}. The system is placed in an arbitrarily oriented
magnetic field and interacts with phonons. 

The system is described by an effective Hamiltonian similar to that introduced in
Ref.~\cite{Gawarecki2018} but generalized to the QDM structure,
\begin{gather}
    H = H_0 + H_{\mathrm{int}},
\end{gather}
where $H_0$ is the effective Hamiltonian of the holes and $H_\mathrm{int}$ stands for hole
interaction with the phonon bath. 

The hole Hamiltonian $H_0$ contains various components accounting for the QD energy
levels, external fields, 
tunnelling and SO coupling \cite{Gawarecki2018}, 
\begin{equation*}
    H_0 = \sum\limits_{\alpha=1}^7 H_0^{(\alpha)}.
\end{equation*}
These components of $H_0$ can be written in the form of a tensor product corresponding to
the decomposition
$\ket{l,m}\otimes \ket{s}$, where the first component refers to the spatial degrees of
freedom and the second one to the spin. The first part,
\begin{equation*}
    H_0^{(\mathrm{1})} = \mathrm{diag} \left( \Delta E_1+e F_z D_z, \Delta E_2, 0, 0
  \right) \otimes \mathbb{I}_2
\end{equation*}
describes the bare energies $\Delta E_1$ and $\Delta E_2$ of the $s$-shell states in the
top and bottom QD, respectively, as well as the effect of the electric field $F_z$ applied
along the stacking direction. The energies are defined with respect to the $p$-shell
energy in the QD2. Next, 
\begin{align*}
    H_0^{(\mathrm{2})} &= \sum\limits_{i={x,y,z}}\frac{1}{2} g_{s,i} \mu_B B_i \\
    &\quad \times \left( \dyad{1,0}{1,0} + \dyad{2,0}{2,0} \right)\otimes \sigma_i 
\end{align*}
and
\begin{align*}
    H_0^{(\mathrm{3})} & =\sum\limits_{i={x,y,z}}\frac{1}{2} g_{p,i} \mu_B B_i \\
    &\quad\times\left( \dyad{2,+1}{2,+1} + \dyad{2,-1}{2,-1} \right)\otimes \sigma_i
\nonumber
\end{align*}
account for Zeeman splittings in the $s$- and $p$-shells, respectively. Anisotropic
L\'ande factors for $s$- and $p$-shell states are denoted by $g_s$ and $g_p$, $\sigma_i$
stand for Pauli matrices, $B_i$ are the components of the magnetic field and $\mu_B$ is the Bohr
magneton. The next term, 
\begin{equation*}
    H_0^{(\mathrm{4})} 
=  V_a \left( \dyad{2,+1}{2,-1} + \dyad{2,-1}{2,+1} \right)\otimes \mathbb{I}_2,
\end{equation*}
describes the influence of an anisotropic QD elongation. Next,
\begin{equation*}
    H_0^{(\mathrm{5})} = 
t_p \left( \dyad{1,0}{2,+1} + \dyad{1,0}{2,-1} + \mathrm{h.c.}\right)\otimes \mathbb{I}_2
\end{equation*}
is the inter-dot $s$-$p$ coupling of strength $t_p$.The part
\begin{equation*}
  H_0^{(\mathrm{6})} = V_{SO} \left( \dyad{2,+1}{2,+1} - \dyad{2,-1}{2,-1} \right) \otimes \sigma_z
\end{equation*}
describes the SO coupling in the $p$ shell. Finally, 
\begin{equation*}
H_0^{(\mathrm{7})} = W B_z \left( \dyad{2,+1}{2,+1} - \dyad{2,-1}{2,-1} \right) \otimes \mathbb{I}_2
\end{equation*}
accounts  for the magnetic effect of the envelope angular momentum.

The hole-phonon interaction Hamiltonian has the form
\begin{gather}\label{eq:H_ph}
    H_{\mathrm{int}} = \sum\limits_{j,j'} \dyad{j}{j'} \sum\limits_{\mathbf{k},\lambda}F_{jj'}^{(\lambda)}\left(\mathbf{k}\right)\left(b_{\mathbf{k}}+b_{-\mathbf{k}}^{\dagger}\right)e^{i\mathbf{kr}},
\end{gather}
where $j$, $j'$ run through all the states in the basis.
Here the phonon annihilation and creation operators of wave vector $\bm{k}$ and
polarization $\lambda$ are denoted by $b_{\bm{k}}$ and $b_{-\bm{k}}^{\dagger}$,
and  
$F_{jj'}^{(\lambda)}\left(\bm{k}\right)$ are hole-phonon coupling constants,
\begin{equation}\label{eq:Fjj'}
  F_{jj'}^{(\lambda)} (\bm{k}) =
    -i\sqrt{\frac{\hbar}{2\rho V c_\lambda k}}\frac{d_p e}{\varepsilon_0\varepsilon_r}
       M_\lambda(\hat{k})\mathcal{F}_{jj'}(k,\hat{k}),
\end{equation}
In the above expression, the wave vector $\bm{k}$ is represented by its length $k$ and
its direction represented in the
spherical coordinates as $\hat{k}=
\left(\sin\theta\cos\varphi,\sin\theta\sin\varphi,\cos\theta \right)$, $\rho$ is the
crystal density, $c_\lambda$ is the sound velocity, 
$d_p$ stands for piezoelectric constant, $\varepsilon_r$ denotes relative electric
permittivity, and
$\mathcal{F}_{jj'}$ is the form-factor 
\begin{align}
    \mathcal{F}_{jj'}\left(\bm{k}\right) = \int \mathrm{d}^3r
  \Psi^*_j(\bm{r})e^{i\bm{kr}}\Psi_{j'}(\bm{r}), 
\end{align}
where $\Psi_j (\bm{r})$ is the envelope wave function of the hole.
In the QDM system the form-factors take the form
\begin{flalign}
    &\mathcal{F}_{jj'}(k, \hat{k}) = f_{jj'}\left(\frac{1}{2}k\sin\theta\right)\nonumber\\
    &\times\exp[-\frac{1}{2}k^2(l_\bot^2\sin^2\theta+l_z^2\cos^2\theta)]\exp\left[i(m_j-m_{j'})\varphi\right]\nonumber\\
    &\times
    \left\{
    \begin{array}{l l}
        \exp\left[ikD_x\sin\theta\cos\varphi+ikD_z\cos\theta\right] & \textrm{for QD1}\\
        1                                                                 & \textrm{for QD2},\\
    \end{array}
    \right.
\end{flalign}
in which $f_{jj'}$ is the matrix
\begin{equation}
    \left\{f_{jj'}\right\}(q) = \left(\begin{array}{cccc}
        1 & 0   & 0     &  0 \\
        0 & 1   & -iq   & -iq\\
        0 & -iq & 1-q^2 & -q^2\\
        0 & -iq & -q^2  & 1-q^2\\
    \end{array}\right)\otimes \mathbb{I}_2
\end{equation}
and $m_j$ is the orbital angular momentum of $j$-th state. The  geometric factor
$M_\lambda$ in Eq.~\eqref{eq:Fjj'} is defined as 
\begin{flalign}\label{eq:M_lambda}
     M_\lambda(\hat{k}) = 2\left[k_x k_y \left(\hat{e}_{\lambda,\bm{k}}\right)_z+k_y k_z \left(\hat{e}_{\lambda,\bm{k}}\right)_x+k_z k_x \left(\hat{e}_{\lambda,\bm{k}}\right)_y\right],
 \end{flalign}
where $\hat{e}_{\lambda,k}$ is the unit polarization vector for the mode ($\lambda, \bm{k}$)

Note that our model does not include the usual SO terms responsible for spin mixing and
spin relaxation within the $s$ shell, so that it allows us to sngle out the effect of
$s$-$p$ coupling and $p$-shell SO structure.

\begin{table}[b]
\centering
\caption{Parameters}
\label{tab:parameters}
\begin{tabular*}{\linewidth}{l @{\extracolsep{\fill}} cc}
\hline
Electric permittivity                   &  $\varepsilon_s$          & $13.2$          \\
Piezoelectric constant                  &  $d$                      & $0.16$ C/m$^2$  \\
Longitudinal sound speed                &  $c_l$                    & $5600$ m/s      \\
Transverse sound speed                  &  $c_t$                    & $2800$ m/s      \\
Density                                 &  $\rho_c$                 & $5360$ kg/m$^3$ \\
Hole wave function widths               &                           &                 \\
\hspace{0.2cm}in-plane                  &  $l_{\bot}$               & $5.0$ nm        \\
\hspace{0.2cm}z direction               &  $l_z$                    & $1.5$ nm        \\
Dots stacking                           &                           &                 \\
\hspace{0.2cm} vertical                 &  $D_z$                    & $7$ nm          \\
\hspace{0.2cm} horizontal               &  $D_x$                    & $1$ nm          \\
Lande g-factor                          &                           &                 \\
\hspace{0.2cm}For s-shell               &                           &                 \\
\hspace{0.4cm}in-plane                  & $g_s^{(x)}, g_s^{(y)} $   & $-0.1$          \\
\hspace{0.4cm}z direction               & $g_s^{(z)}$               & $-5.51$         \\
\hspace{0.2cm}For p-shell               &                           &                 \\
\hspace{0.4cm}in-plane                  & $g_p^{(x)}, g_p^{(y)} $   & $0.05$          \\
\hspace{0.4cm}z direction               & $g_p^{(z)}$               & $2.62$          \\
SO coupling                     & $V_{SO}$                  & $-9.79$ meV     \\
Envelope angular \\momentum coefficient   & $W$                       & $-0.532$ meV/T  \\
Inter-dot s-p coupling                  &  $t_p$                    & $0.1$ meV       \\
Intra-dot s-p energy separation         & $\Delta E_2$              & $-40$ meV       \\
Inter-dot s-p energy separation         & $\Delta E_1$              & $-7$ meV        \\
\hline
\end{tabular*}
  \label{tab:shape-functions}
\end{table}

The values of the parameters can be found in Tab.~\ref{tab:parameters}. The value of the hole
$s$-$p$ tunnel coupling, which is a crucial parameter in the presented calculations, is
assumed to be one order of magnitude lower than the analogous value calculated for the
electron \cite{Gawarecki2014}. This seems to be a safe estimate in view of the \kp results
that yield similar orders of magnitude for the electron \cite{gawarecki10} and hole
\cite{gawarecki12a} $s$-shell tunnel couplings.

\subsection{Spin relaxation rate}

We find the spin-flip rate between the $s$-shell Zeeman sublevels of the
ground hole states in the QD1 using the Fermi golden rule.
We diagonalize effective Hamiltonian to find the eigenstates 
\begin{equation}\label{eigen}
    \ket{\alpha} = \sum_j c_{\alpha j}\ket{j}
\end{equation}
and the transition frequencies $\omega_{if}=\left(E_i-E_f\right)/ \hbar$, where initial
and final states are the nominally spin-up and spin-down $s$-shell states in the top QD. 
We define the spectral densities (in terms of the basis states)
\begin{align*}
   \lefteqn{R_{jj'nn'}(\omega) =}\\
&\quad \sum\limits_{\lambda=l,t_1,t_2} 
\int d\Omega_{\hat{k}} \dfrac{1}{c_\lambda^3}M_\lambda^2(\hat{k})
F_{jj'}\left(\frac{\omega}{c_\lambda},\hat{k}\right)
F^{*}_{nn'}\left(\frac{\omega}{c_\lambda},\hat{k}\right),
\end{align*}
where $l,t_{1},t_{2}$ denote the longitudinal and transverse phonon branches.
Using the Fermi golden rule and writing the system eigenstates as a superposition of the
basis states according to Eq.~\eqref{eigen}, 
we obtain a formula for the spin relaxation rate, which is a combination of the spectral
densities taken at the transition frequency 
\begin{equation}\label{eq:rate_final}
\gamma_{\alpha\rightarrow\alpha'} =\sum_{j,j', n n'} c_{\alpha j'}^* c_{\alpha' j}
c_{\alpha n'} c_{\alpha' n}^*R_{j j' nn'}(\omega_{if}).
\end{equation}

\section{Results}\label{sec:results}

\begin{figure}[tb]
    \centering
    \includegraphics[width=\columnwidth]{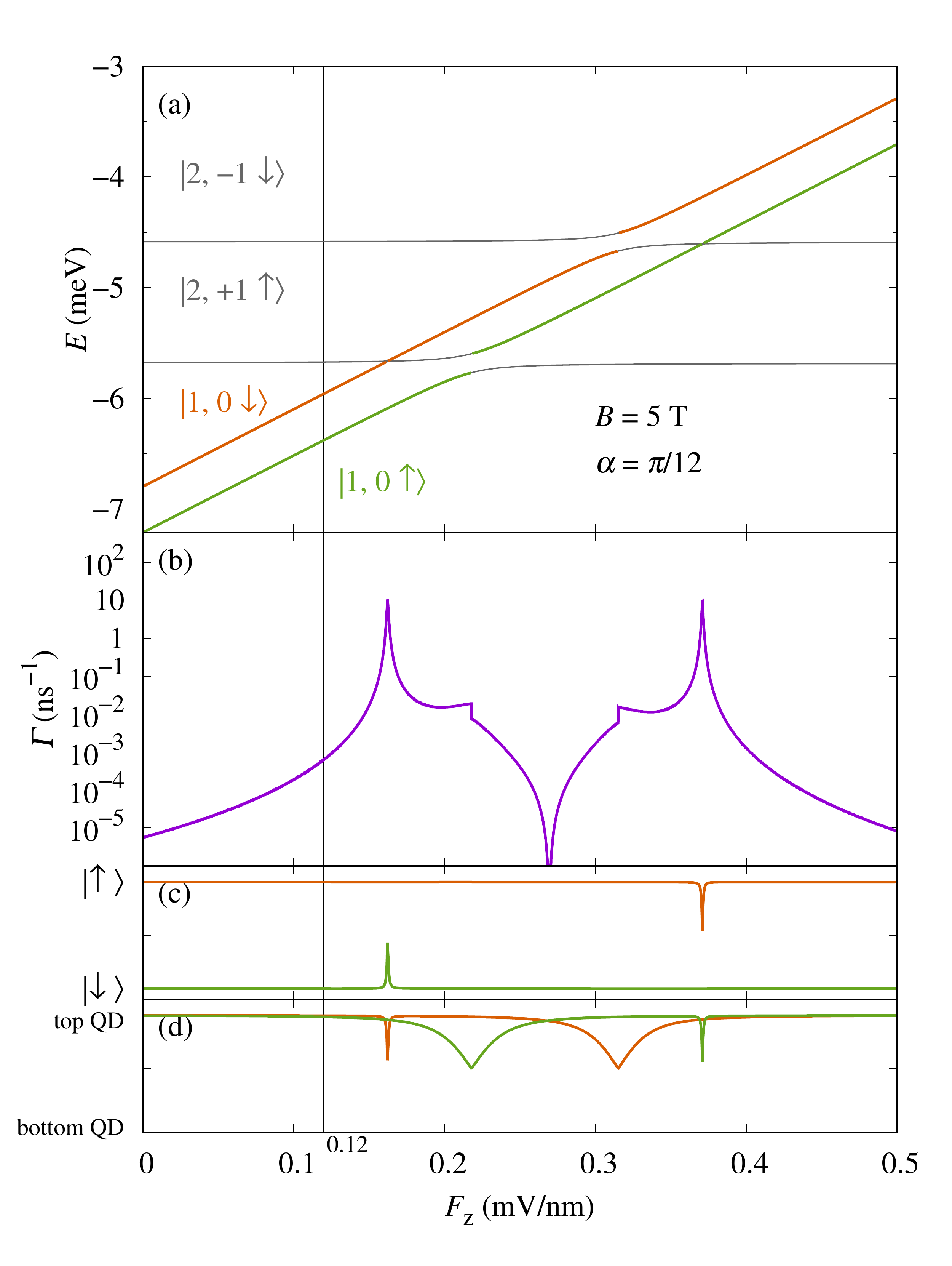}
    \vspace{-0.5cm}\caption{(a) The eigenstates of the system in the energy range around
      the $s$-$p$ resonance; colors mark the states localized in QD1. (b) Hole spin
      relaxation rate between the Zeeman sublevels in the $s$ shell of QD1. (c) Spin
      polarization of the two states. (d) Localization of the two states. Colors in (c)
      and (d) refer to the branches marked in (a). The vertical line marks the value used
      in Fig.~\ref{fig:states_magn}.} 
    \label{fig:states_elec} 
\end{figure}

The rate of phonon-induced spin relaxation in the QD1 caused by the coupling to the
$p$-shell of the QD2 was obtained by numerical implementation of the model discussed in
Section~\ref{sec:Model}. 
Fig.~\ref{fig:states_elec}(a) shows the diagram of energy levels as a function of electric
field for the range of field magnitudes around the resonances between the $s$ shell of QD1
and the $p$ shell of QD2. The results were obtained for the magnetic field of $5$~T, 
tilted by $15^{\circ}$ off the plane. The
states localized in QD1, distinguishable by their evolution 
in the electric field, are marked with colors. The two branches of the $s$ states are the
Zeeman sublevels with respect to the spin quantization axis determined by the field
orientation. The two $p$-shell states are the lower half of the four $p$-shell 
states split by the SO interaction \cite{Gawarecki2018}, corresponding to
anti-parallel orientation of orbital and band angular momenta (the upper two states lie above
the axis range). 

Fig.~\ref{fig:states_elec}(b) shows the transition rate between the $s$-shell Zeeman
sublevels of QD1 (orange to green in Fig.~\ref{fig:states_elec}(a)) calculated from
Eq.~\eqref{eq:rate_final}. 
The rate peaks at the resonances with the $p$-shell states, where the life time of the upper
state is reduced to about 100 ps. In order to interpret this, we present the spin
polarization of the two states in Fig.~\ref{fig:states_elec}(c) (with color coding
referring to Fig.~\ref{fig:states_elec}(a)) and their localization in
Fig.~\ref{fig:states_elec}(d). At the resonance, one of the Zeeman branches in QD1 mixes
with a state with the opposite spin from the $p$ shell of QD2. This results in a reduction
of the spin polarization of this state from $\pm 1$ to 0, as can be seen in
Fig.~\ref{fig:states_elec}(c) and delocalization of the state between the QDs, visible in
Fig.~\ref{fig:states_elec}(d). As a result, the spin-flip time becomes comparable to the
charge relaxation time (tens of picoseconds). 

This resonant enhancement is, however, constrained to a very narrow parameter range, since
the resonance between states with opposite spins is very narrow (the corresponding
anticrossing is not even visible in  Fig.~\ref{fig:states_elec}(a)). 
Away from the resonance the rate decreases, as expected for an effect that
relies on the admixture of $p$ states. Still, in a range of a few meV from the resonance,
the spin-flip rates remain relatively high and correspond to spin life times in the $\mu$s
range, which are shorter than those observed in some experiments. This shows that in a
certain range of energies around the resonance with the $p$ 
shell of the QD2 the hole spin flip in the ground state of the QD1 is dominated by the
inter-QD $s$-$p$ coupling and $p$-shell SO effects.

\begin{figure}[tb]
    \centering
    \includegraphics[angle=270,width=\linewidth]{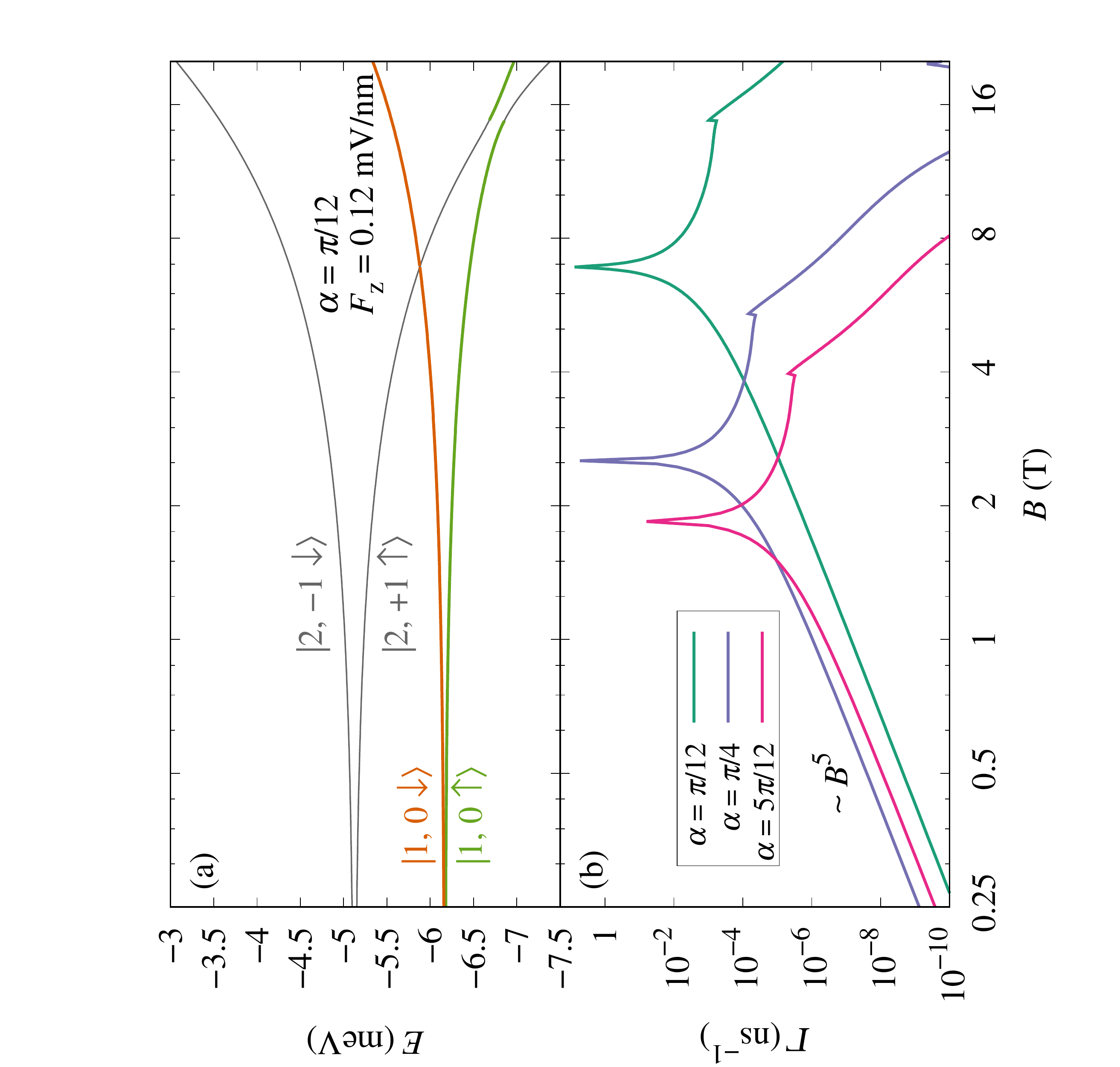}
    \vspace{-0.5cm}\caption{The eigenstate diagram (a) and hole spin relaxation rate (b)
      as a function of magnetic field for $F_{z}=0.12$~mV/nm and tilt angles (from the
      sample plane) as shown.} 
    \label{fig:states_magn}
\end{figure}

\begin{figure}[tp]
    \centering
    \includegraphics[angle=270,width=\linewidth]{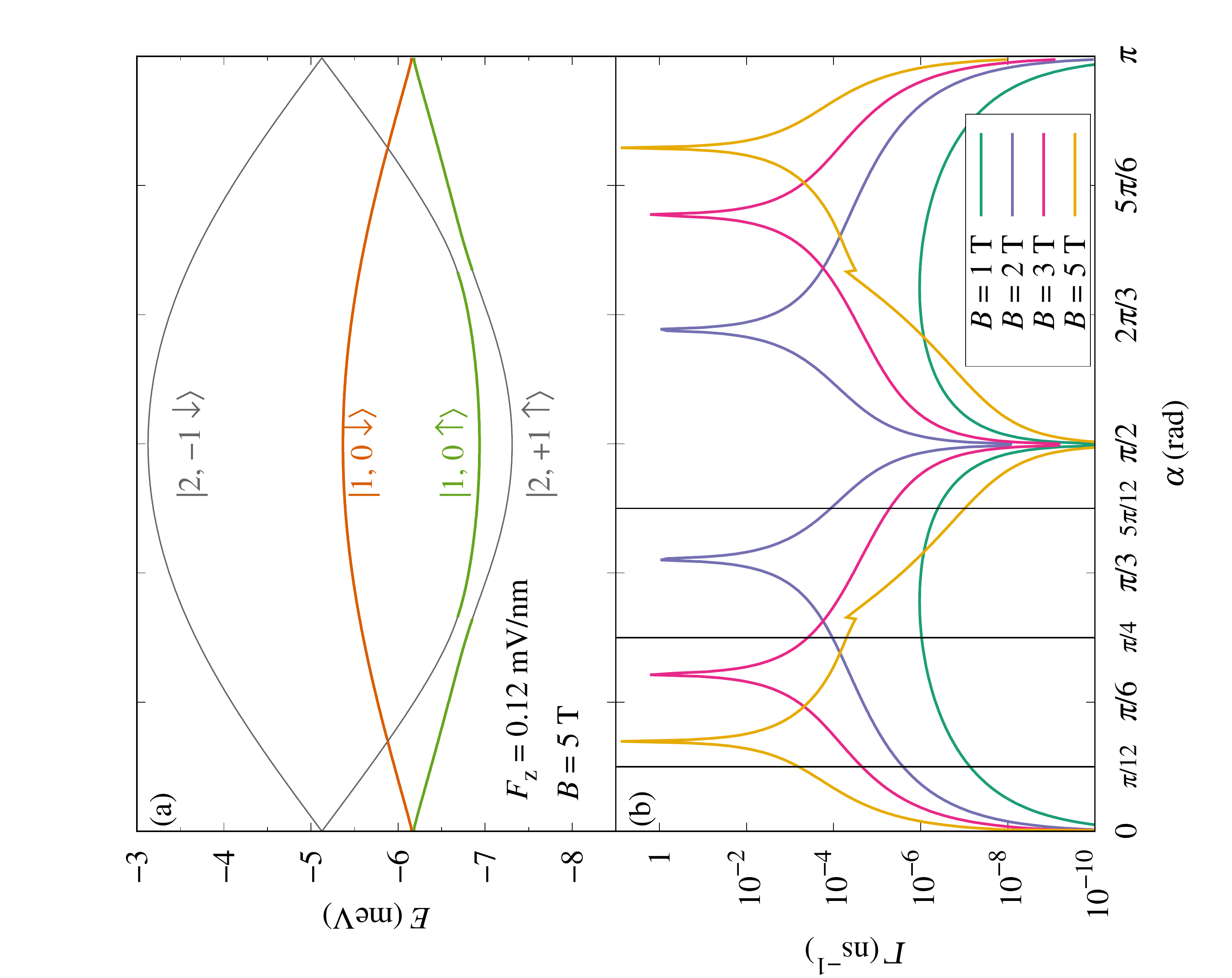}
    \vspace{-0.5cm}\caption{The eigenstate diagram (a) and the hole spin relaxation rate
      (b) as a function of the orientation of magnetic field (angle with respect to the
      in-plane direction) for magnetic field magnitudes as shown. The
      electric field has the magnitude of $0.12$~mV/nm.} 
    \label{fig:angle} 
\end{figure}

In Fig~\ref{fig:states_magn} we show the system spectrum and relaxation rates as a
function of magnetic field for a fixed electric field of $0.12$~mV/nm (marked by a
vertical line in Fig.~\ref{fig:states_elec}.). The rates in  Fig.~\ref{fig:states_magn}(b)
are plotted for three different tilt angles (marked by vertical lines in
Fig.~\ref{fig:angle}) and the spectrum in
Fig.~\ref{fig:states_magn}(a) is shown for one angle 
$\pi/12$, corresponding to the green line in  Fig~\ref{fig:states_magn}(b). 
The rates show a peak at the resonance, the origin of which is the same as in the previous
discussion. Away from the resonances, the rate grows as $B^5$. 

In Fig.~\ref{fig:angle} the eigenstate diagram and the spin relaxation rate are presented
as a function of the magnetic field orientation for different magnetic field magnitudes. 
The rates in  Fig~\ref{fig:angle}(b)
are plotted for various magnitudes of the magnetic field and the spectrum in
Fig~\ref{fig:angle}(a) corresponds to $B=5$~T (yellow line in  Fig~\ref{fig:angle}(b)). 
Apart from the resonant enhancements, one can see considerable reduction of the rate at
angles 0 and $\pi/2$ (exact Voigt and Faraday geometry, respectively). The former is due
to the strong anisotropy of the hole g-factor and nearly vanishing $s$-shell Zeeman
splitting in the Voigt geometry, which suppresses the phonon-assisted relaxation due to
low phonon spectral density at low frequencies. In the Faraday geometry the $z$ projection
of the spin is strictly conserved in our model, as follows directly from the form of the
Hamiltonian.

\section{Conclusions}
\label{sec:concl}

We have studied phonon-assisted hole-spin relaxation rate between Zeeman sublevels in a QD
forming part of a QDM. We have calculated the energy states using an effective Hamiltonian
that accounts for the spin-orbit coupling of the other QD in the QDM and tunnel coupling
between the $s$ and $p$ shells of the two dots. We have shown that the inter-QD $s$-$p$
coupling induced by QD misalignment combined with strong SO effects in the $p$ shell
leads to enhanced spin relaxation in the vicinity of the $s$-$p$ resonance, with the
exception of exact Voigt and Faraday geometries, when the rates becomes small or vanish,
respectively. The typical 
spin relaxation times resulting from the $s$-$p$ coupling mechanism are on the order of 
$\mu$s even away from the resonance, when the states are nearly completely localized and
spin-polarized. The mechanism studied here can therefore dominate over other known
spin-flip processes in a certain energy interval around the $s$-$p$ resonance. 

The authors acknowledge funding from the Polish National Science Centre (NCN) under Grant
No.~2016/23/G/ST3/04324 (KK) and Grant No.~2014/13/B/ST3/04603 (PM).

\bibliography{library}
\end{document}